\documentclass[12pt]{article}
\usepackage{amssymb,amsmath,mathrsfs}
\usepackage{graphicx,epsfig}
\usepackage{caption}
\usepackage{subcaption}
\usepackage{enumerate}
\usepackage{epstopdf}
\usepackage{xcolor}
\usepackage{float}
\usepackage{rotating}
\usepackage{tikz}
\usepackage{dsfont}
\usepackage{comment} 
\usepackage{tabularx}
\usepackage{hyperref}
\usepackage{upgreek}
\usepackage[margin=3cm]{geometry}
\usepackage[english]{babel}
\usepackage[square,numbers]{natbib} 
\usepackage[linesnumbered,lined,boxed,commentsnumbered]{algorithm2e}
\RestyleAlgo{ruled}
\SetKwComment{Comment}{/* }{ */}
\hypersetup{
    colorlinks=true,
    linkcolor=cyan,
    filecolor=cyan,      
    urlcolor=cyan,
    citecolor = cyan
    }
\graphicspath{{Rips sequences/}}
 \numberwithin{equation}{section}

\DeclareUnicodeCharacter{0342}{\~{n}}
\DeclareUnicodeCharacter{0342}{\'{A}}

\begin{document}
\newcounter{eqnarray}
\newtheorem{t1}{Theorem}[section]
\newtheorem{d1}{Definition}[section]
\newtheorem{c1}{Corollary}[section]
\newtheorem{l1}{Lemma}[section] \newtheorem{r1}{Remark}[section]
\newtheorem{e1}{Example}[section]
\newtheorem{p1}{Proposition}[section]
\newtheorem{a1}{Result}[section]

\newcommand{\calH}{\mathcal{H}}
\newcommand{\inpr}[2]{\left\langle #1, #2 \right\rangle_\calH}

\title{Operator on Operator Regression in Quantum Probability}
\author{\small 
	Suprio Bhar \\
	\small IIT Kanpur\\
	\small Department 
 of Mathematics and Statistics \\
	\small  Kanpur 208016, India\\
	{\small Email: suprio@iitk.ac.in }\\
	\and
	\small Subhra Sankar Dhar \\
	\small  IIT Kanpur\\
	\small   Department of Mathematics and Statistics \\
	\small Kanpur 208106, India\\
	{\small Email: subhra@iitk.ac.in}\\
        \and
        \small Soumalya Joardar\\
        \small IISER Kolkata\\
        \small Department of Mathematics and Statistics\\
        \small Mohanpur 741246, India\\
        {\small Email : Soumalya@iiserkol.ac.in}
}
\maketitle
\begin{center}
    \textbf{Abstarct} 
\end{center} 

This article introduces operator on operator regression in quantum probability. Here in the regression model, the response and the independent variables are certain operator valued observables, and they are linearly associated with unknown scalar coefficient (denoted by $\beta$), and the error is a random operator. In the course of this study, we propose a quantum version of a  class of estimators (denoted by $M$ estimator) of $\beta$, and the large sample behaviour of those quantum version of the estimators are derived, given the fact that the true model is also linear and the samples are observed eigenvalue pairs of the operator valued observables. 

\section{Introduction}
Regression analysis is one of the most well-known toolkit in statistical modelling. More specifically, regression analysis studies the ``relationship" between the dependent/response variable and one or more independent/explanatory variables. This ``relationship" can be linear, non-linear or even non-parametric as well (see, e.g., \cite{MR4511147} and a few relevant references therein), and using various statistical techniques, one may estimate the unknown parameters involved in the ``relationship". Recently, the concepts of regression have been extended to functional valued random elements also; for instance, the interested reader may look at \cite{MR4665579} and a few relevant references therein regarding function on function regression. However, to the best of our knowledge, regression analysis is an area unexplored in the setting of quantum probability. This work aims to investigate various statistical issues involved in operator on operator regression in quantum probability. We first provide a literature review and then brief the research problem without much technical details in the following. 

In the context of other statistical concepts studied in the language of quantum probability theory, recently, \cite{MR4534161} extended the concepts of sufficient statistic and Rao-Blackwellization in the framework of quantum probability, and a few years before, \cite{MR2256497} overviewed and reviewed the problem of statistical sufficiency in the setting of quantum statistics. They provided sufficient and necessary conditions for sufficiency and established many results from traditional statistical theory in the language of noncommutative geometry, and at the same time, \cite{MR2207329} derived the noncommutative version of well-known factorization theorem in traditional Statistics (see, e.g, \cite{MR2002723}). 

Let us now consider a pair of operator valued observables $(X, Y)$ (see, e.g., \cite{MR0013657} for details on observables) with observed $n$ pairs of eigen values $(\lambda_1, \mu_1), \ldots, (\lambda_{n}, \mu_n)$. Observe that here in the setting of quantum probability, we are observing the eigen values of $(X, Y)$, unlike the observations on the random element $(X, Y)$ as in traditional Statistics/classical Probability. Suppose that unobserved operators $Y$ and $X$ are linearly associated in true sense with scalar valued coefficient (see \eqref{modeltrue}), and in the sense of regression modelling, the operator valued observables are assumed to be linear with scalar valued coefficient having some operator valued error (see \eqref{model}). In this work, based on $(\lambda_1, \mu_1), \ldots, (\lambda_{n}, \mu_n)$, we estimate (denoted by $\hat{\beta}_{n}$ as an estimator) the unknown parameter $\beta$ involved in the model described in \eqref{model}. In the course of this study, we establish the connection between the concepts in quantum probability and classical probability (see Section \ref{quantumclassical}), and in view of this relation, we investigate the large sample properties $\hat{\beta}_{n}$ (see Section \ref{LSP}).  

The rest of the article is structured as follows. Section \ref{PM} describes the preliminaries of the model in terms of quantum probability, and for this problem, a key connection between the quantum probability and classical probability is investigated in Section \ref{quantumclassical}, and the problem is reformulated in terms of classical probability in Section \ref{refor}. The large sample properties of the proposed estimator is studied in Section \ref{LSP}, and Section \ref{CR} contains a few relevant concluding remarks. Finally, necessary technical details are provided in the Appendix in Section \ref{TD}.

\section{Preliminaries : Model}\label{PM}

Let $\calH$ be a real separable Hilbert space with inner-product $\inpr{\cdot}{\cdot}$ and norm $\|\cdot\|_\calH$. If $T$ and $S$ are two bounded linear operators on $\calH$ related by $T = \beta S$ for some non-zero real scalar $\beta$, then given any eigen-value $\lambda$ of $S$, we have an eigen-value $\mu = \beta \lambda$ of $T$ with the same eigen vector of unit norm. Again, given any eigen-value $\mu$ of $T$, we have an eigen-value $\lambda = \frac{\mu}{\beta}$ of $S$ with the same eigen vector of unit norm. In light of this fact, one may expect that the linear regression model associating two operator valued observables can be represented by their corresponding eigen values. The research problem is foramlly described in the following.  

Let $X$ and $Y$ be two compact self-adjoint operator (on $\calH$) valued observables related by 
\begin{eqnarray}\label{modeltrue}
Y = \beta_{0} X, 
\end{eqnarray} where $\beta_{0}\in\mathbb{R}$ is unknown. We now consider the operator on operator regression model
\begin{eqnarray}\label{model}
Y = \beta X + \epsilon,  
\end{eqnarray} where $\epsilon$ is a random operator valued error, and $\beta\in\mathbb{R}$ is an unknown constant. In this work, our objective is to propose an estimator (denoted by $\hat{\beta}_{n}$) of the unknown $\beta$ based on the observed eigenvalues of $(X, Y)$. 
The technical assumptions are stated in Section \ref{LSP}, where the large sample properties of $\hat{\beta}_{n}$ are studied.

Suppose that $(\lambda_1, \mu_1), \ldots, (\lambda_n, \mu_n)$ are $n$ pairs of observed eigen-values of $(X, Y)$. Motivated by our earlier observation on deterministic operators $T$ and $S$, we assume that each pair $(\lambda_j, \mu_j)$ appear with a common eigen-vector $v_j$ of unit norm. Here, as mentioned earlier, using the observed $(\lambda_{j}, \mu_{j})_{j = 1}^{n}$, we would like to estimate the unknown parameter $\beta$. The estimation procedure of $\beta$ is described in the following. 

In order to carry our the estimation of $\beta$, we first reformulate the problem, from originally stated in terms of $X$ and $Y$ in the quantum probability setup to their eigen-value pairs $(\lambda, \mu)$ in the classical probability setup. Next, using the classical probability model thus obtained the problem is reduced to a standard linear regression problem and we can then use classical statistical tools in order to estimate $\beta$. Section \ref{quantumclassical} studies the eigen decomposition of operator valued observables $X$ and $Y$, which is the key concept in reformulating the problem in the setup of classical probability. 

\subsection{Distribution of eigen decomposition of $X$ and $Y$ : From quantum to classical probability}\label{quantumclassical}

Assume that $X$ and $Y$ are compact self-adjoint operators on $\calH$ with discrete spectrum given by $\sigma(X)$ and $\sigma(Y)$, respectively. As per our discussion above, the eigenspaces corresponding to each feasible eigen-value pairs are the same. Consequently, we have the following spectral decompositions (see \cite{KnappBk}):
\[X = \sum_{\alpha \in \sigma(X)} \alpha P_\alpha, \quad Y = \sum_{\alpha \in \sigma(Y)} \alpha P_\alpha,\]
where the sum is countable, and $P_\alpha$ denotes the projection onto the eigen-space of any eigen-value $\alpha$. Here, the eigenvalues are all real and the eigen-spaces corresponding to $0 \neq \alpha \in \sigma(X)$ are finite-dimensional. Now, given any state $\phi$ (unit trace), we have the expectations in the setting of quantum probability as

\[E X = tr(\phi X) = \sum_{\alpha \in \sigma(X)} \alpha \, tr(\phi P_\alpha), \quad E Y = tr(\phi Y) = \sum_{\alpha \in \sigma(Y)} \alpha \, tr (\phi P_\alpha).\]

Note that $tr(\phi P_\alpha) \geq 0, \forall \alpha \in\sigma(X)$. Moreover, $\sum_{\alpha \in \sigma(X)} tr(\phi P_\alpha) = tr(\rho) = 1$. A similar statement is true when $X$ is replaced by $Y$. 

\begin{r1}\label{eigendistribution}
Observe that one can think of $\lambda_j$'s drawn independently from a discrete probability distribution supported on $\sigma(X)$ and with the probability mass function $$f_\lambda(\alpha) = tr(\phi P_\alpha) 1_{(\alpha \in \sigma(X))}.$$ Similarly, $\mu_j$'s are drawn independently from a discrete probability distribution supported on $\sigma(Y)$ and with the probability mass function $$f_\mu(\alpha) = tr(\phi P_\alpha)1_{(\alpha \in \sigma(Y))}.$$ This fact enables us to reformulate the problem in the set up of traditional statistics.
\end{r1}

\subsection{A reformulation : estimation of $\beta$}\label{refor}

Given $n$ pairs $(\lambda_1, \mu_1), \ldots, (\lambda_n, \mu_n)$ of observed eigen-values of $(X, Y)$, assume that $v_1, v_2, \cdots, v_n$ be the corresponding eigen-vectors of unit norm. Then,
\begin{eqnarray}\label{eigenX}
X v_{j} = \lambda_{j} v_{j}~\mbox{for all $j = 1, \ldots, n$}, 
\end{eqnarray} where for $j = 1, \ldots, n$,  $||v_{j}||_\calH = 1$, and similarly, we have 
\begin{eqnarray}\label{eigenY}
Y v_{j} = \mu_{j} v_{j}~\mbox{for all $j = 1, \ldots, n$}. 
\end{eqnarray}
Next, it follows from \eqref{model}, \eqref{eigenX} and \eqref{eigenY} that for all $j = 1, \ldots, n$, $Y v_{j} = \beta X v_{j} + \epsilon v_{j}$, which yields
\begin{equation}
\mu_{j}v_{j} = \beta \lambda_{j}v_{j} + \epsilon v_{j},
\end{equation}
and consequently,
\begin{equation}\label{reduced-model}
\mu_j = \inpr{v_j}{\mu_jv_j} = \inpr{v_j}{\beta \lambda_{j}v_{j} + \epsilon v_{j}} = \beta \lambda_j + \inpr{v_j}{\epsilon v_j},\, \forall j = 1, \ldots, n.
\end{equation}

We now estimate $\beta$, involved in \eqref{reduced-model}, based on $(\lambda_{j}, \mu_{j})_{j = 1}^{n}$ in various ways, by assuming conditions on the error terms $\inpr{v_j}{\epsilon v_j},\, j = 1, \ldots, n$. Now, recall Remark \ref{eigendistribution} that $\lambda_j$'s are i.i.d. random variables following a discrete probability distribution supported on $\sigma(X)$ and with the probability mass function $$f_\lambda(\alpha) = tr(\phi P_\alpha) 1_{(\alpha \in \sigma(X))},$$ and $\mu_j$'s are i.i.d. random variables following a discrete probability distribution supported on $\sigma(Y)$ and with the probability mass function $$f_\mu(\alpha) = tr(\phi P_\alpha)1_{(\alpha \in \sigma(Y))}.$$

 For the data $(\lambda_{j}, \mu_{j})$ ($j = 1, \ldots, n$), the $M$-estimator of $\beta$ (see \eqref{reduced-model}), which is denoted by $\hat{\beta}_{n}$, is defined as 
\begin{eqnarray}\label{Mestimation}
\hat{\beta}_{n} &=& arg\min_{\beta}\sum\limits_{j = 1}^{n}  \rho(\inpr{v_j}{\epsilon v_j})\nonumber\\
& = & arg\min_{\beta}\sum\limits_{j = 1}^{n} \rho(\mu_{j} - \beta \lambda_{j})~\mbox{(uisng \eqref{reduced-model})}, 
\end{eqnarray} where $\rho(.)$ is a certain convex function, and few more assumptions will be stated in the subsequent section. Observe that for $\rho(x) = x^{2}$, $\hat{\beta}_{n}$ coincides with the well-known least squares estimator (see, e.g., \cite{MR1652247}, pp.\ 42--43) based on the data $(\lambda_{j}, \mu_{j})_{j = 1}^{n}$, and when $\rho (x) = |x|$, $\hat{\beta}_{n}$ coincides with the well-known least absolute deviation estimator (see, e.g., \cite{MR1652247}, pp.\ 42--43) based on the data $(\lambda_{j}, \mu_{j})_{j = 1}^{n}$. Besides, it also includes Huber's estimator with $\rho(x) = x^{2}1_{(|x|\leq c)}/2 + (c|x| - c^{2}/2)1_{(|x| > c)}$, where $c> 0$ (see, e.g., \cite{MR1652247}, p.\ 43), ${\cal{L}}^{q}$ regression estimate with $\rho(x) = |x|^{q}$, where $q\in [1, 2]$ (see, e.g., \cite{4c2d121a-4dfe-33c1-a5b9-56d247d33a74}) and well-known $\alpha$-th regression quantiles with $\rho(x) = |x| + (2\alpha - 1) x$, where $\alpha\in (0, 1)$ (see, e.g., \cite{MR1652247}, p.\ 43, equation (5.5)).  This is one of the reasons to consider the technique of $M$-estimation, which includes many well-known estimators of the unknown regression parameter. 

\section{Large sample properties of $\hat{\beta}_{n}$}\label{LSP}

In order to study the large sample properties of $\hat{\beta}_{n}$ defined in \eqref{Mestimation},  one needs to assume the following technical conditions. 

\noindent (A1) $\rho(.)$ is a strictly convex function. 

\noindent (A2) The error random variables $\inpr{v_j}{\epsilon v_j}$ ($j = 1, \ldots, n$) are i.i.d., and the common distribution of $\inpr{v_j}{\epsilon v_j}$, which is denoted by $F$ (depends on joint distribution of $(\lambda, \mu)$), is such that $F ({\cal{D}}) = 0$. Here ${\cal{D}}$ is the set of discontinuity points of $\rho'(.)$, where $\rho{'}(.)$ denotes the derivative of $\rho(.)$.

\noindent (A3) $E[\rho{'} (\inpr{v_1}{\epsilon v_1} + c)] = ac + o(|c|)$ as $|c|\rightarrow 0$. Here $a$ and $c$ are two some constants. 

\noindent (A4) $g(c) : = E[\rho{'} (\inpr{v_1}{\epsilon v_1} + c) - \rho{'} (\inpr{v_1}{\epsilon v_1})]$ exists for some $c\in (-\delta, \delta)$, where $\delta > 0$ is a some constant. Moreover, $g(.)$ is a continuous function at $0$.

\noindent (A5) $E[\{\rho{'} (\inpr{v_1}{\epsilon v_1})\}^{2}] : = D > 0$.

\noindent (A6) $S_{n} : = \sum\limits_{j = 1}^{n} \lambda_{j}^{2} > 0$ for some $n\geq n_{0}$ and 
$$d_{n}^{2} : = \max_{1\leq j \leq n}\frac{\lambda_{j}^{2}}{S_{n}}\rightarrow 0\mbox~~{as}~~n\rightarrow\infty .$$

\begin{r1}\label{dis}
Condition (A1) is required to have unique minima of the equation described in \eqref{Mestimation}, and it is a common assumption across the literature of M-estimation (see, e.g., \cite{MR1652247}) in traditional Statistics. Condition (A2) indicates certain smoothness of the function $\rho$ along with the fact that the error terms (in the forms of inner product) are i.i.d. random variables. Conditions (A3) and (A4) imply the existence of the derivative of the criterion function of $M$ estimation in some neighbourhood of $0$, which is necessary to solve minimization problem involved in $M$ estimation. Next, Condition (A5) imply that the $M$-estimator $\hat{\beta}$ (see \eqref{Mestimation}), after appropriate normalization, will weakly converge to a non-degenerate random variable (see the statement of Theorem \ref{an} below, and finally, Condition (A6) is required to have the asymptotic normality of $\hat{\beta}_{n}$, and from statistical point of view, one can interpret as the variation explained by any eigen value will not be a dominating factor.
\end{r1}

Now, we state the consistency and the asymptotic normality results associated with 
$\hat{\beta}_{n}$.

\begin{t1}\label{consistency}
Under (A1)--(A6), $\hat{\beta}_{n}\stackrel{p}\rightarrow \beta$ as $n\rightarrow\infty$. Here 
$\stackrel{p}\rightarrow$ denotes the convergence in probability, and $\hat{\beta}_{n}$ and $\beta$ are the same as defined in \eqref{Mestimation} and \eqref{reduced-model}, respectively.  
\end{t1}

\begin{t1}\label{an}
Under (A1)--(A6), $$T_{n}^{-1/2}K_{n} (\hat{\beta}_{n} - \beta)\stackrel{w}\rightarrow Z,$$ where $Z$ is a random variable associated with standard normal distribution, and $\stackrel{w}\rightarrow$ denotes the convergence in distribution. Here $T_{n} = D S_{n}$ and $K_{n} = a S_{n}$, where $D$, $a$ and $S_{n}$ are the same as defined in (A5), (A3) and (A6), respectively. 
\end{t1}

\begin{r1}
In the set up of quantum probability, the observations $(\lambda_{j}, \mu_{j})_{j = 1}^{n}$ are $n$ eigen value pairs of $(X, Y)$ unlike the classical probability, where the observations are $(X_{j}, Y_{j})_{j = 1}^{n}$ obtained from the distribution of $(X, Y)$. Despite this issue, Theorems \ref{consistency} and \ref{an} assert that $\hat{\beta}_{n}$ is a consistent estimator of $\beta$, and after appropriate normalization, $(\hat{\beta}_{n} - \beta)$ converges weakly to a standard normal random variable. 
\end{r1}

\section{Concluding Remarks}\label{CR}
In this work, observe that the true model (see \eqref{modeltrue}) is also linear with scalar valued coefficient, i.e., in other words, we work on specified model. However, the study will be entirely different if the model becomes mis-specified, which is of great interest in many statistical problem (see, e.g., \cite{MR4751814} and a few relevant references therein). Investigating this issue may be of interest to future research. Besides, we have assumed our operators to be compact and self-adjoint. The classical analog would be to assume that the real random variables are discrete. It will be of interest when the involved operators are just self-adjoint. Then the spectrum may be continuous and in this case, our technique would not suffice. Moreover, in this work as the parameter $\beta$ is scalar, the operators involved here are apriori simultaneously disgonalizable and therefore they can be measured simultaneously.  


In the course of this study, it is assumed that the error random variables $\inpr{v_j}{\epsilon v_j}$ ($j = 1, \ldots, n$) involved in \eqref{reduced-model} are i.i.d., and the assertions in the main results (i.e., Theorems \ref{consistency} and \ref{an}) rely on this assumption. However, this assumption can be weakened by following the approach considered in \cite{10.1214/009053606000001406}. In this work, we did not dig in this issue as establishing the large sample properties of $\hat{\beta}_{n}$ is not the main theme of the work, and therefore, we kept the assumption simpler so that the readers can appreciate the notion on quantum probability conveniently for operator on operator regression.  

\noindent{\bf Acknowledgement:} Subhra Sankar Dhar gratefully acknowledges his core research grant (CRG/2022/001489), Government of India. Suprio Bhar gratefully acknowledges the Matrics grant MTR/2021/000517 from the Science and Engineering Research Board (Department of Science \& Technology, Government of India). Soumalya Joardar acknowledges support from the SERB MATRICS grant (MTR 2022/000515).

\bibliographystyle{apalike} 
\bibliography{main}

\begin{thebibliography}{}

\bibitem[Bagchi and Dhar, 2024]{MR4751814}
Bagchi, P. and Dhar, S.~S. (2024).
\newblock Characterization of the least squares estimator: mis-specified multivariate isotonic regression model with dependent errors.
\newblock {\em Theory Probab. Math. Statist.}, (110):143--158.

\bibitem[Bai et~al., 1992]{7c660722-2e21-32e3-8b4b-5803dfa9d2bd}
Bai, Z.~D., Rao, C.~R., and Wu, Y. (1992).
\newblock M-estimation of multivariate linear regression parameters under a convex discrepancy function.
\newblock {\em Statistica Sinica}, 2(1):237--254.

\bibitem[Dette and Tang, 2024]{MR4665579}
Dette, H. and Tang, J. (2024).
\newblock Statistical inference for function-on-function linear regression.
\newblock {\em Bernoulli}, 30(1):304--331.

\bibitem[Dhar et~al., 2022]{MR4511147}
Dhar, S.~S., Jha, P., and Rakshit, P. (2022).
\newblock The trimmed mean in non-parametric regression function estimation.
\newblock {\em Theory Probab. Math. Statist.}, (107):133--158.

\bibitem[Dirac, 1945]{MR0013657}
Dirac, P. A.~M. (1945).
\newblock On the analogy between classical and quantum mechanics.
\newblock {\em Rev. Modern Phys.}, 17:195--199.

\bibitem[Jen\v~cov\'a and Petz, 2006a]{MR2207329}
Jen\v~cov\'a, A. and Petz, D. (2006a).
\newblock Sufficiency in quantum statistical inference.
\newblock {\em Comm. Math. Phys.}, 263(1):259--276.

\bibitem[Jen\v~cov\'a and Petz, 2006b]{MR2256497}
Jen\v~cov\'a, A. and Petz, D. (2006b).
\newblock Sufficiency in quantum statistical inference. {A} survey with examples.
\newblock {\em Infin. Dimens. Anal. Quantum Probab. Relat. Top.}, 9(3):331--351.

\bibitem[Knapp, 2005]{KnappBk}
Knapp, A.~W. (2005).
\newblock {\em Advanced real analysis}.
\newblock Cornerstones. Birkh\"auser Boston, Inc., Boston, MA.
\newblock Along with a companion volume {\it Basic real analysis}.

\bibitem[Lai and Lee, 2005]{4c2d121a-4dfe-33c1-a5b9-56d247d33a74}
Lai, P.~Y. and Lee, S. M.~S. (2005).
\newblock An overview of asymptotic properties of lp regression under general classes of error distributions.
\newblock {\em Journal of the American Statistical Association}, 100(470):446--458.

\bibitem[Shao, 2003]{MR2002723}
Shao, J. (2003).
\newblock {\em Mathematical statistics}.
\newblock Springer Texts in Statistics. Springer-Verlag, New York, second edition.

\bibitem[Sinha, 2022]{MR4534161}
Sinha, K.~B. (2022).
\newblock Sufficient statistic and {R}ao-{B}lackwell theorem in quantum probability.
\newblock {\em Infin. Dimens. Anal. Quantum Probab. Relat. Top.}, 25(4):Paper No. 2240005, 16.

\bibitem[van~der Vaart, 1998]{MR1652247}
van~der Vaart, A.~W. (1998).
\newblock {\em Asymptotic statistics}, volume~3 of {\em Cambridge Series in Statistical and Probabilistic Mathematics}.
\newblock Cambridge University Press, Cambridge.

\bibitem[Wu, 2007]{10.1214/009053606000001406}
Wu, W.~B. (2007).
\newblock {M-estimation of linear models with dependent errors}.
\newblock {\em The Annals of Statistics}, 35(2):495 -- 521.

\end{thebibliography}

\section{Appendix : Technical Details}\label{TD}

\noindent {\bf Proof of Theorem \ref{consistency}:} The arguments in the proof are parallel to the proof of Theorem 2.2 in \cite{7c660722-2e21-32e3-8b4b-5803dfa9d2bd}. Without loss of generality, we assume that $\beta = 0$. It is now enough to show that 
$$P[|\hat{\beta}_{n}| \geq a_{n}]\rightarrow 0$$ as $n\rightarrow\infty$, where $\{a_{n}\}_{n\geq 1}$ is such that $a_{n}\rightarrow\infty$ as $n\rightarrow\infty$. Observe that it follows from (3.11) in \cite{7c660722-2e21-32e3-8b4b-5803dfa9d2bd} that there exists a sequence $a_{n}^{'}$ such that $a_{n}^{'}\rightarrow\infty$ as $n\rightarrow\infty$ and $a_{n}^{'}\leq a_{n}$ for all $n\in\mathbb{N}$. Moreover, in view of it, we also have 
\begin{eqnarray}\label{ineq1}
\sup_{|\beta|\leq a_{n}^{'}}\left|\sum\limits_{j = 1}^{n}[\rho(\inpr{v_j}{\epsilon v_j} - \beta\lambda_{j}) - \rho(\inpr{v_j}{\epsilon v_j}) + \beta\lambda_{j}\rho{'}(\inpr{v_j}{\epsilon v_j})] - \frac{\beta^{2}K_{n}}{2} \right|\stackrel{p}\rightarrow 0
\end{eqnarray} as $n\rightarrow\infty$, where $K_{n}$ is the same as defined in the statement of Theorem \ref{an}. 

Further, when $|\beta| = a_{n}^{'}$ for some $n\in\mathbb{N}$, then observe that 
\begin{eqnarray}\label{ineq2}
\frac{\beta^{2}K_{n}}{2}\geq\frac{a_{n}^{'2}\max\limits_{1\leq j\leq n}\lambda_{j}}{2}, 
\end{eqnarray} which follows from the definition of of $K_{n}$ defined in the statement of Theorem \ref{an} and $S_{n}$ defined in (A6) before Remark \ref{dis}. Moreover, observe that using (A4) on \eqref{ineq1}, we have 
\begin{eqnarray}\label{ineq3}
\sum\limits_{j = 1}^{n}\lambda_{j} \rho{'}(\inpr{v_j}{\epsilon v_j}) = O_{p}(1) ,    
\end{eqnarray} and hence, 
\begin{eqnarray}\label{ineq4}
\sup_{|\beta|\leq a_{n}^{'}}\left|\beta\sum\limits_{j = 1}^{n}\lambda_{j} \rho{'}(\inpr{v_j}{\epsilon v_j})\right| = O_{p}(a_{n}^{'}).  
\end{eqnarray}

Therefore, using \eqref{ineq1}, \eqref{ineq2}, \eqref{ineq3} and \eqref{ineq4}, we have 
\begin{eqnarray}\label{ineq5}
 P\left(\inf_{|\beta| = a_{n}^{'}}\sum\limits_{j = 1}^{n}\rho(\inpr{v_j}{\epsilon v_j} - \beta\lambda_{j}) - \rho(\inpr{v_j}{\epsilon v_j}) \leq 0\right)\rightarrow 0  
\end{eqnarray} as $n\rightarrow\infty$, and next, using (A1), we have 
\begin{eqnarray}\label{ineq6}
 P\left(\inf_{|\beta| \geq a_{n}^{'}}\sum\limits_{j = 1}^{n}\rho(\inpr{v_j}{\epsilon v_j} - \beta\lambda_{j}) - \rho(\inpr{v_j}{\epsilon v_j}) \leq 0\right)\rightarrow 0  
\end{eqnarray} as $n\rightarrow\infty$. 

Finally, by the definition of $\hat{\beta}_{n}$ in \eqref{Mestimation}, we have 
\begin{eqnarray}\label{ineq7}
P(|\hat{\beta}_{n}|\geq a_{n}^{'})\rightarrow 0    
\end{eqnarray} as $n\rightarrow\infty$. As $a_{n}^{'}\leq a_{n}$ for all $n\in\mathbb{N}$, the proof follows. \hfill$\Box$

\noindent {\bf Proof of Theorem \ref{an}:} Without loss of generality, we here consider $\beta = 0$. First of all, using (A6) and in view of Lindeberg Feller CLT (see \cite{MR1652247}), we have 
\begin{eqnarray}\label{inweakcon}
\sum\limits_{j = 1}^{n} \lambda_{j} \rho{'}(\inpr{v_j}{\epsilon v_j}) : = K_{n}\widetilde{\beta}_{n}\stackrel{w}\rightarrow Z,  \end{eqnarray} as $n\rightarrow\infty$, where $Z$ is a random variable associated with standard normal distribution, $K_{n}$ is the same as defined in the statement of Theorem \ref{an}, and `$\stackrel{w}\rightarrow$' denotes weak convergence.
 Hence, in order to prove this theorem, it is enough to show that 
\begin{eqnarray}\label{rqconprob}
\hat{\beta}_{n} -\widetilde{\beta}_{n}\stackrel{p}\rightarrow 0
\end{eqnarray} as $n\rightarrow\infty$, 
where $\widetilde{\beta}_{n}$ and $\hat{\beta}_{n}$ are the same as defined in \eqref{inweakcon} and \eqref{Mestimation}, respectively. Observe that using \eqref{ineq1} and the definition of $\widetilde{\beta}_{n}$ in \eqref{inweakcon}, we 
have 
\begin{eqnarray}\label{ineq7}
 \sum\limits_{j = 1}^{n}[\rho(\inpr{v_j}{\epsilon v_j} - \widetilde{\beta}_{n}\lambda_{j}) - \rho(\inpr{v_j}{\epsilon v_j})] + \frac{\beta\widetilde{\beta}_{n} K_{n}}{2}\stackrel{p}\rightarrow 0   
\end{eqnarray} as $n\rightarrow\infty$, and for some sequence $\{b_{n}\}_{n\geq 1}$ such that $b_{n}\rightarrow\infty$ as $n\rightarrow\infty$ and for some arbitrary constant $\delta > 0$, we have 
\begin{eqnarray}\label{ineq8}
\sup_{|\beta|\leq b_{n} + \delta} \left|\sum\limits_{j = 1}^{n}[\rho(\inpr{v_j}{\epsilon v_j} - \beta\lambda_{j}) - \rho(\inpr{v_j}{\epsilon v_j})] + \beta\widetilde{\beta}_{n} K_{n} - \frac{\beta^{2}K_{n}}{2}\right|\stackrel{p}\rightarrow 0   
\end{eqnarray} as $n\rightarrow\infty$. Now, it follows from \eqref{ineq7} and \eqref{ineq8}, we have 
\begin{eqnarray}\label{ineq9}
\sup_{|\beta - \widetilde{\beta}_{n}| = \delta} \left|\sum\limits_{j = 1}^{n}[\rho(\inpr{v_j}{\epsilon v_j} - \beta\lambda_{j}) - \rho(\inpr{v_j}{\epsilon v_j} - \widetilde{\beta}_{n}\lambda_{j})]  - \frac{(\beta - \widetilde{\beta}_{n})^{2}K_{n}}{2}\right|\stackrel{p}\rightarrow 0     
\end{eqnarray} as $n\rightarrow\infty$. Moreover, observe that when $|\beta - \widetilde{\beta}_{n}| = \delta$, we have 
\begin{eqnarray}\label{ineq10}
(\beta - \widetilde{\beta}_{n})^{2}K_{n}\geq\left(\min_{1\leq j \leq n}\lambda_{j}\right)\delta^{2}. 
\end{eqnarray} Therefore, using \eqref{ineq9} and \eqref{ineq10}, we have 
\begin{eqnarray}
P(|\beta - \widetilde{\beta}_{n}|> \delta)\rightarrow 0    
\end{eqnarray} as $n\rightarrow\infty$, where $\delta > 0$ is an arbitrary constant. It completes the proof. \hfill$\Box$
\end{document}